\documentclass[12pt,dvips]{article}

\usepackage{cite,epsfig}
\usepackage{color}
\usepackage{axodraw}

\def\theequation{\arabic{section}.\arabic{equation}}

\begin{document}

\begin{flushright}
MAN/HEP/2006/39\\
hep-ph/0612188 \\
December 2006
\end{flushright}

\begin{center}
{\bf {\Large Radiative Yukawa Couplings for Supersymmetric\\[3mm]
    Higgs Singlets at Large {\boldmath $\tan \beta$} }}
\end{center}

\bigskip\bigskip

\begin{center}
{\large Robert N. Hodgkinson and Apostolos Pilaftsis}\\[3mm]
{\em School of Physics and Astronomy, University of Manchester}\\
{\em Manchester M13 9PL, United Kingdom}
\end{center}

\vspace{2cm}

\centerline{\bf ABSTRACT}
\noindent
Singlet  Higgs bosons  present  in  extensions of  the  MSSM can  have
sizable Yukawa  couplings to the $b$  quark and the  $\tau$ lepton for
large  values of  $\tan\beta$  at  the 1-loop  level.   We present  an
effective  Lagrangian  which  incorporates these  $\tan\beta$-enhanced
Yukawa couplings and which enables us to study their effect on singlet
Higgs-boson phenomenology within the context of both the mnSSM and the
NMSSM.  In particular, we find that the loop-induced coupling can be a
significant effect for the  singlet pseudoscalar, and may dominate its
decay modes.  Further  implications of the $\tan\beta$-enhanced Yukawa
couplings  for  the phenomenology  of  the  singlet  Higgs bosons  are
briefly discussed.

\newpage

\section{Introduction}

The  Minimal Supersymmetric  extension  of the  Standard Model  (MSSM)
provides  a  self-consistent  framework  to  technically  address  the
gauge-hierarchy  problem  related to  the  Standard  Model (SM)  Higgs
boson.   The   2-Higgs  doublet  potential  of  the   MSSM  is  highly
constrained  at  the  tree  level  which  makes  it  very  predictive.
Radiative   corrections   from   third   generation   squarks   affect
significantly the mass  of the lightest neutral Higgs  boson $H_1$ and
generically shift  it upwards by  an amount of $\sim  30$--40~GeV. The
$H_1$ boson has  to be lighter than about 135~GeV  for large values of
$\tan\beta \stackrel{>}{{}_\sim}  20$, where $\tan\beta$  is the ratio
of the two Higgs vacuum  expectation values (VEVs).  For low values of
$\tan\beta \sim 3$, most of the  parameter space that gives a mass for
the lightest  neutral Higgs boson  above the present  LEP experimental
limits has  been excluded.  For a comprehensive  analysis performed by
the LEP Higgs working group, see~\cite{ADLO}.

One   theoretical   weakness   of    the   MSSM   is   the   so-called
$\mu$-problem~\cite{Kim:1983dt,CPNSH}.   This is  related to  the fact
that  the  $\mu$-parameter  describing  the  mixing  of  the  2  Higgs
superfields    in   the   superpotential,    i.e.~$\mu   \widehat{H}_u
\widehat{H}_d$,  has  to be  of  order  the  soft SUSY-breaking  scale
$M_{\rm  SUSY}  \sim   1$~TeV,  for  successful  electroweak  symmetry
breaking.  Instead,  within the  context of supergravity  (SUGRA), the
$\mu$-parameter is in general not  protected by gravity effects and so
expected to be of order  Planck mass $M_{\rm Pl}$.  A natural solution
to the $\mu$-problem may be  obtained in extensions of the MSSM, where
the $\mu$-term  has been promoted  to a dynamical variable,  e.g.~to a
SM-singlet  chiral superfield  $\widehat{S}$. In  such a  setting, the
scalar component  $S$ of $\widehat{S}$  generically acquires a  VEV of
order  $M_{\rm SUSY}$,  thereby giving  rise  to a  $\mu$-term of  the
required order.

Since     the     $\mu$-term    is     replaced     by    the     term
$\lambda\widehat{S}\widehat{H}_u\widehat{H}_d$  in  singlet extensions
of  the MSSM, the  resulting superpotential  will exhibit  an unwanted
global  Peccei--Quinn (PQ)  symmetry $U(1)_{\rm  PQ}$,  unless further
additions or assumptions are made  to the model.  The PQ symmetry must
be   explicitly   broken   to   avoid  the   appearance   of   visible
electroweak-scale axions  after the spontaneous  symmetry breaking SSB
of $U(1)_{\rm PQ}$. The choice  of discrete and gauged symmetries used
to  break the  PQ symmetry  distinguish between  the  different models
which have been studied  in the literature~\cite{CPNSH}, including the
Next-to-Minimal  Supersymmetric  SM  (NMSSM)\cite{NMSSM}, the  minimal
nonminimal  Supersymmetric  SM (mnSSM)\cite{Panagiotakopoulos:2000wp},
the $U(1)'$-extended Supersymmetric SM (UMSSM)\cite{Cvetic:1997ky} and
the       secluded       $U(1)'$-extended      Supersymmetric       SM
(sMSSM)\cite{Erler:2002pr}.

Irrespectively of  the details of the particular  model, singlet Higgs
bosons have  no tree level couplings  to SM fermions  or gauge bosons.
It has long  been known~\cite{TBanks,Hempfling:1993kv} that within the
MSSM, threshold corrections to the  Yukawa couplings to $b$ quarks and
$\tau$  leptons   can  become  significant  in  the   limit  of  large
$\tan\beta$, which partially overcome the loop suppression factor.  In
regions where  the mixing between  the Higgs particles  is negligible,
the  one-loop correction  can  dominate the  $H_1\rightarrow b\bar  b$
decay width \cite{Coarasa:1995yg}.

In this paper we show  that an analogous $\tan\beta$ enhancement takes
place for the Yukawa couplings  of the singlet Higgs bosons in minimal
extensions of  the MSSM. In particular, we  explicitly demonstrate how
the effective  couplings $Sb\bar b$ and  $S\tau^+\tau^-$ are generated
radiatively  through  squark-gaugino  loops  and  their  size  can  be
significant, e.g.~of order  the SM Yukawa couplings.  In  the limit in
which the $H_d$ doublet decouples  from the low energy spectrum, these
one-loop  couplings provide  the  dominant decay  mechanism for  light
singlets.   Recent work has  proposed the  possibility of  light Higgs
singlets   as   a   solution   to   the   little   hierarchy   problem
\cite{Dermisek:2005ar}.   In this  scenario the  threshold corrections
are not only important for the Higgs searches themselves, but they may
potentially provide  a first signpost towards  the physically realized
region of SUSY parameter space.

The  paper  is organized  as  follows: in  section  2  we present  the
effective Lagrangian for the Higgs-boson couplings to $b\bar b$ and to
$\tau^+\tau^-$,  along  with  analytic  expressions for  the  dominant
contributions. Section  3 discusses the  phenomenological implications
of  the radiative  singlet-Higgs Yukawa  couplings for  the  mnSSM and
NMSSM. Our notations and conventions regarding the Higgs sector follow
those of the first paper in~\cite{Panagiotakopoulos:2000wp}, whilst 
those regarding the squarks and charginos are outlined in Appendix~A.
Our conclusions and possible future directions are presented in Section~4.



\setcounter{equation}{0}
\section{Effective Yukawa Couplings}
\label{effectivelagrangian}

In  this  section,  we  derive  the  general  form  of  the  effective
Lagrangian  for   the  self-energy   transition  $f_L  \to   f_R$,  in
CP-conserving  Higgs singlet  extensions of  the MSSM,  where  $f$ may
represent a $b$-quark or a $\tau$-lepton.  We then use the Higgs-boson
Low Energy  Theorem (HLET)  \cite{Ellis:1975ap,BPP} to compute  the 1-loop
effective Higgs-boson couplings to $b$ quarks and $\tau$ leptons.

The general  effective Lagrangian for the  self-energy transition $f_L
\to  f_R$ in  the  background  of non-vanishing  Higgs  fields may  be
written down as
\begin{eqnarray}
-\,\mathcal{L}^f_{\rm self} &=& h_f\; \bar{f}_R
\,\bigg(\, \Phi^{0\ast}_1\: +\: \Delta_f [\Phi^0_1,\Phi^0_2,S]\, \bigg)
\,f_L\ +\ {\rm h.c.},
\label{lagrangianone}
\end{eqnarray} 
where   $\Phi^0_{1,2} = \frac{1}{\sqrt{2}}\left(v_{1,2} + \phi_{1,2} +
ia_{1,2}\right)$ are the electrically neutral components of the two Higgs
doublets $H_{d,u}$\footnote{Here we adopt the convention for the Higgs
doublets: $H_u  \equiv \Phi_2$,  $H_d \equiv i\tau_2  \Phi^*_1$, where
$\tau_2$  is the  usual Pauli  matrix.}, $S = \frac{1}{\sqrt{2}}\left(
v_S + \phi_S + i a_S\right)$ is the singlet Higgs  field and  the  functional
$\Delta_f[\Phi^0_1,\Phi^0_2,S]$  encodes  the  radiative  corrections.
Given that  the VEV  of the effective  Lagrangian $-\mathcal{L}^f_{\rm
self}$  should equal  the fermion  mass $m_f$,  an expression  for the
effective Yukawa coupling $h_f$ can be found, i.e.
\begin{equation}
h_f\ =\ \frac{g_w m_f}{\sqrt{2} M_W c_\beta}\; 
\left(1+{\sqrt{2}\over v_1}\left<\Delta_f\right>\right)^{-1}\,,
\end{equation} 
where  $\left<\Delta_f\right>$ is  the  VEV of  $\Delta_f$, where  the
renormalization scale  is set at $M_{\rm SUSY}$.  The contributions to
the functional $\Delta_f$ which  get enhanced at large $t_\beta \equiv
\tan\beta$ have been well understood  within the framework of the MSSM
\cite{Hempfling:1993kv,Coarasa:1995yg}.      As     is    shown     in
Fig~\ref{gauginos}, the dominant contributions to $\Delta_f$ come from
diagrams with  gluinos and  bottom squarks and  with chargino  and top
squarks in the loop.

We may  relate the  self-energy effective Lagrangian  ${\cal L}^f_{\rm
self}$ to  the effective Lagrangian  for the Higgs-boson  couplings to
the fermion $f$,  using the HLET~\cite{BPP}. In terms  of the physical
Higgs  fields  $H_{1,2,3}$ and  $A_{1,2}$,  the effective  interaction
Lagrangian reads:
\begin{equation}
  \label{Leff}
-\,\mathcal{L}^{\rm eff}_{\phi\bar ff}\ =\ 
{g_wm_f\over 2M_W}\; \left[\; \sum_{i=1}^3g^S_{H_iff}H_i\bar
ff\ +\ \sum_{i=1}^2 g^P_{A_iff}A_i
\left(\bar fi\gamma^5 f\right)\, \right]\; , 
\end{equation} 
with
\begin{eqnarray}
  \label{gS}
g^S_{H_i ff} &=& \left(1+{\sqrt{2}\over v_1} 
\left<\Delta_f\right>\right)^{-1}\,
\left[\, {O^H_{1i}\over c_\beta}
+ \Delta_f^{\phi_2}{O^H_{2i}\over c_\beta} +
\Delta_f^{\phi_S}{O^H_{3i}\over c_\beta}\, \right]\, ,\\
  \label{gA}
g^P_{A_i ff} &=& \left(1+{\sqrt{2}\over v_1}
\left<\Delta_f\right>\right)^{-1}\,
\left[\, -\left(t_\beta+\Delta_f^{a_2}\right)O^A_{1i}
+\Delta_f^{a_S}{O^A_{2i}\over c_\beta}\, \right]\; .  
\end{eqnarray} 
Here the orthogonal matrix $O^H~(O^A)$ is related to the mixing of the
CP-even (CP-odd)  scalars and  the loop corrections  are given  by the
HLET
\begin{equation}
\Delta_f^{\phi_{2,S}}\ =\
\sqrt{2}\; \left<{\partial\Delta_f\over\partial\phi_{2,S}}\right>\;,\qquad
\Delta_f^{a_{2,S}}\ =\
i\,\sqrt{2}\; \left<{\partial\Delta_f\over\partial a_{2,S}}\right>\; .
\end{equation} 
In~(\ref{gS})   and~(\ref{gA}),  we  have   neglected  the   one  loop
contributions  to  the  $\phi_1$   coupling,  since  they  are  small,
i.e.~$\Delta_f^{\phi_1}\ll\Delta_f^{\phi_{2,S}}t_\beta$.

\subsection{Effective $b$-quark Yukawa Couplings}
\label{byukawacouplings}

\begin{figure}
\begin{center}
  \includegraphics{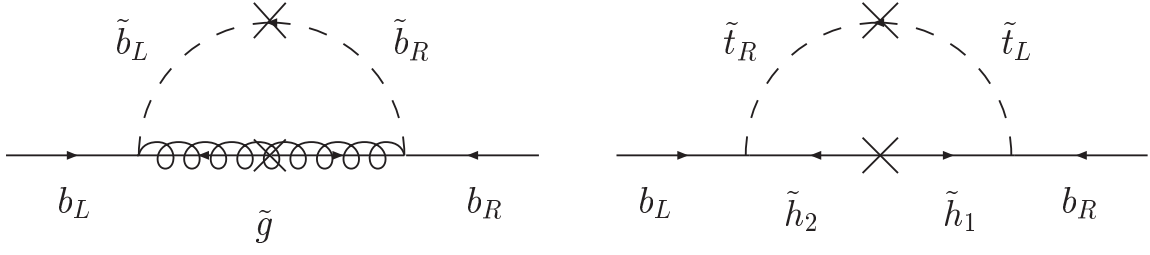}
\caption{\it Dominant contributions to the bottom quark self energy at 
large $t_\beta$, in the single VEV insertion approximation.} \label{gauginos}
\end{center}
\end{figure}

As  in the  MSSM, there  are $t_\beta$-enhanced  contributions  to the
self-energy of  the $b$ quark  from both gluino and  chargino exchange
diagrams,   as   shown   in  Fig.~\ref{gauginos}.   Evaluating   these
$t_\beta$-enhanced diagrams  at zero external  momentum 
and neglecting subdominant terms proportional to $\alpha_w$ yields
\begin{eqnarray}
\Delta_b & = &
-\; {2\alpha_s\over 3\pi}\; M_3\,
    \left(A_b\Phi^{0\ast}_1-\lambda S^\ast\Phi^{0\ast}_2\right)
I(m^2_{\tilde b_1},m^2_{\tilde b_2},M^2_3) \nonumber \\
&& +\ {{h_t^2}\over 16\pi^2}\; 
\left(A_t\Phi_2^{0\ast}-\lambda S\Phi_1^{0\ast}
\right)\ \left[\ m_{\tilde \chi_1} 
{\mathcal V}^\dag_{21}
{\mathcal U}^\ast_{12}\; 
I(m^2_{\tilde t_1}, m^2_{\tilde t_2},
 m^2_{\tilde\chi_1})\right. \nonumber \\
&&\ \left. +\ m_{\tilde \chi_2}
{\mathcal V}^\dag_{22}
{\mathcal U}^\ast_{22}\; 
I(m^2_{\tilde t_1}, m^2_{\tilde t_2},
 m^2_{\tilde\chi_2})\; \right]
\end{eqnarray}
In the above, $I(a,b,c)$  
is the usual 1-loop integral function given by
\begin{eqnarray}
I(a,b,c)\ =\ {{ab\ln{(a/b)}\: +\: bc\ln{(b/c)}\: +\: ac\ln{(c/a)}}\over
(a-b)(b-c)(a-c)}\; ,
\end{eqnarray}
and  ${\mathcal V},\;  {\mathcal U}$  are  the  chargino-mixing 
matrices  defined  in
Appendix~A.   Note   that  ${\mathcal V},\;  {\mathcal U}$  are 
  functionals  of
$\Phi^0_{1,2}$   and   $S$,   as   are  the   sbottom   quark   masses
$m_{\tilde{b}_{1,2}}$,  stop  quark  masses $m_{\tilde{t}_{1,2}}$  and
chargino masses $m_{\tilde{\chi}_{1,2}}$. Explicit expressions for
the masses and mixing angles are given in Appendix~A.


\begin{figure}
\begin{center}
 \includegraphics[scale=0.9]{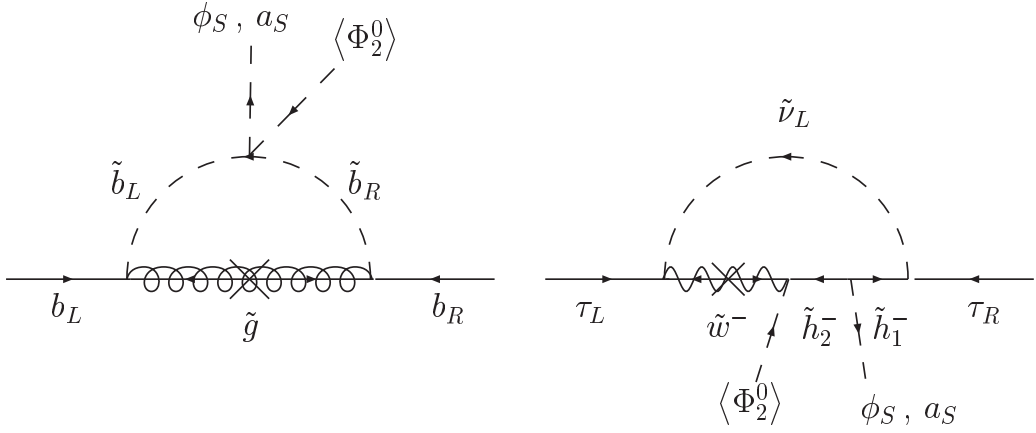}
\caption{\it Dominant contributions to the couplings $\phi_S\bar bb$,
$\phi_S\tau^+\tau^-$, $a_S\bar bb$ and $a_S\tau^+\tau^-$, in the large
$t_\beta$ limit.}\label{domcoup}
\end{center}
\end{figure}

We may then use the HLET to calculate the corresponding graphs with an
additional zero  momentum Higgs  insertion.  The lowest  order $\phi_S
b\bar  b$ graphs  are  shown in  Fig.~\ref{domcoup}. Neglecting
again terms proportional to $\alpha_w$, the coupling parameters
$\Delta_b^{\phi_{2,S}}$ are given by
\begin{eqnarray}
  \label{delta2}
\Delta_b^{\phi_2} &=& {2\alpha_s \over
    3\pi}\; M_3\; \Bigg[\,\mu\; I(m^2_{\tilde b_1},m^2_{\tilde
    b_2},M^2_3)\ -X_b v_1{\partial\over\partial\phi_2}\ 
I(m^2_{\tilde  b_1},m^2_{\tilde b_2},M^2_3)\,\Bigg]\nonumber\\
&& +\ {h_t^2\over 16\pi^2}\; A_t\; \Bigg[\ m_{\tilde\chi_1} 
{\mathcal V}^\dag_{21}
{\mathcal U}^\ast_{12}\; 
I(m^2_{\tilde t_1}, m^2_{\tilde t_2}, m^2_{\tilde\chi_1})
+m_{\tilde\chi_2}
{\mathcal V}^\dag_{22}
{\mathcal U}^\ast_{22}\; 
I(m^2_{\tilde t_1}, m^2_{\tilde t_2},
 m^2_{\tilde\chi_2}) \Bigg]\ \\
&&+\ {h_t^2 \over 16\pi^2}\; X_t v_2\;
{\partial\over\partial\phi_2} \Bigg[\ m_{\tilde\chi_1} 
{\mathcal V}^\dag_{21}
{\mathcal U}^\ast_{12}\; 
I(m^2_{\tilde t_1}, m^2_{\tilde t_2}, m^2_{\tilde\chi_1})
+m_{\tilde\chi_2}
{\mathcal V}^\dag_{22}
{\mathcal U}^\ast_{22}\; 
I(m^2_{\tilde t_1}, m^2_{\tilde t_2},
 m^2_{\tilde\chi_2}) \Bigg]\; ,\nonumber\\[3mm]
  \label{deltaS}
\Delta_b^{\phi_S} &=&
{2\alpha_s\over 3\pi}\; M_3 \,
\Bigg[\,\mu\;{v_2\over v_S}
 I(m^2_{\tilde b_1},m^2_{\tilde b_2},M^2_3)\ 
-X_b v_1{\partial\over\partial\phi_S}\ 
I(m^2_{\tilde b_1},m^2_{\tilde b_2},M^2_3)\, \Bigg]\nonumber\\
&&-\ {h_t^2\over 16\pi^2}\; \mu {v_1\over v_S}\; \Bigg[\ m_{\tilde\chi_1} 
{\mathcal V}^\dag_{21}
{\mathcal U}^\ast_{12}\; 
I(m^2_{\tilde t_1}, m^2_{\tilde t_2}, m^2_{\tilde\chi_1})
+m_{\tilde\chi_2}
{\mathcal V}^\dag_{22}
{\mathcal U}^\ast_{22}\; 
I(m^2_{\tilde t_1}, m^2_{\tilde t_2},
 m^2_{\tilde\chi_2}) \Bigg]\\
&& +\ {h_t^2 \over 16\pi^2}\; X_t v_2\;
{\partial\over\partial\phi_S} \Bigg[\ m_{\tilde\chi_1} 
{\mathcal V}^\dag_{21}
{\mathcal U}^\ast_{12}\; 
I(m^2_{\tilde t_1}, m^2_{\tilde t_2}, m^2_{\tilde\chi_1})
+m_{\tilde\chi_2}
{\mathcal V}^\dag_{22}
{\mathcal U}^\ast_{22}\; 
I(m^2_{\tilde t_1}, m^2_{\tilde t_2},
 m^2_{\tilde\chi_2}) \Bigg] \; ,\nonumber
\end{eqnarray}
where $X_b  = A_b - \mu t_\beta$  and $X_t = A_t  - \mu/t_\beta$.  The
derivatives  act on  all Higgs-dependent  functionals to  their right,
generating  a   rather  lengthy  expression  which  we   do  not  show
here explicitly.  These  derivative terms represent higher  number of Higgs
insertions,  beyond  the usual  single  Higgs insertion  approximation
often followed  in the literature.  They can  be $t_\beta$-enhanced in
certain regions  of the parameter  space, especially for  the chargino
case,  and  are  therefore  consistently  included  in  our  numerical
analysis in Section~3.

The presence of the singlet in the model does not alter the 1-loop
$\tan\beta$ enhanced couplings of the doublet Higgs fields well-known
 from the MSSM \cite{CGNW}. As a consistency check, we have compared 
the gluino-exchange terms of (\ref{delta2}) with corresponding results
 from the first reference of \cite{CGNW}, and we find that the HLET
calculation gives identical results.


Correspondingly,     the      loop-induced     coupling     parameters
$\Delta_b^{a_{2,S}}$ are given by
\begin{eqnarray}
  \label{Dba2}
\Delta_b^{a_2} &=& {2\alpha_s \over
    3\pi}\; M_3\; \,\mu\; 
I(m^2_{\tilde  b_1},m^2_{\tilde b_2},M^2_3)\, \nonumber\\
&& +\ {h_t^2\over 16\pi^2}\; A_t\; \Bigg[\ m_{\tilde\chi_1} 
{\mathcal V}^\dag_{21}
{\mathcal U}^\ast_{12}\; 
I(m^2_{\tilde t_1}, m^2_{\tilde t_2}, m^2_{\tilde\chi_1})
+m_{\tilde\chi_2}
{\mathcal V}^\dag_{22}
{\mathcal U}^\ast_{22}\; 
I(m^2_{\tilde t_1}, m^2_{\tilde t_2},
 m^2_{\tilde\chi_2}) \Bigg]\ \\
&&+\ {h_t^2 \over 16\pi^2}\; X_t v_2\;
{\partial\over\partial a_2} \Bigg[\ m_{\tilde\chi_1} 
{\mathcal V}^\dag_{21}
{\mathcal U}^\ast_{12}\; 
I(m^2_{\tilde t_1}, m^2_{\tilde t_2}, m^2_{\tilde\chi_1})
+m_{\tilde\chi_2}
{\mathcal V}^\dag_{22}
{\mathcal U}^\ast_{22}\; 
I(m^2_{\tilde t_1}, m^2_{\tilde t_2},
 m^2_{\tilde\chi_2}) \Bigg]\; ,\nonumber\\[3mm]
  \label{DbaS}
\Delta_b^{a_S} &=&
{2\alpha_s\over 3\pi}\; M_3 \,
\,\mu\;{v_2\over v_S}
 I(m^2_{\tilde b_1},m^2_{\tilde b_2},M^2_3)\nonumber\\
&&+\ {h_t^2\over 16\pi^2}\; \mu {v_1\over v_S}\; \Bigg[\ m_{\tilde\chi_1} 
{\mathcal V}^\dag_{21}
{\mathcal U}^\ast_{12}\; 
I(m^2_{\tilde t_1}, m^2_{\tilde t_2}, m^2_{\tilde\chi_1})
+m_{\tilde\chi_2}
{\mathcal V}^\dag_{22}
{\mathcal U}^\ast_{22}\; 
I(m^2_{\tilde t_1}, m^2_{\tilde t_2},
 m^2_{\tilde\chi_2}) \Bigg] \\
&& +\ {h_t^2 \over 16\pi^2}\; X_t v_2\;
{\partial\over\partial a_S} \Bigg[\ m_{\tilde\chi_1} 
{\mathcal V}^\dag_{21}
{\mathcal U}^\ast_{12}\; 
I(m^2_{\tilde t_1}, m^2_{\tilde t_2}, m^2_{\tilde\chi_1})
+m_{\tilde\chi_2}
{\mathcal V}^\dag_{22}
{\mathcal U}^\ast_{22}\; 
I(m^2_{\tilde t_1}, m^2_{\tilde t_2},
 m^2_{\tilde\chi_2}) \Bigg] \; ,\nonumber
\end{eqnarray}
Notice  that both  $\Phi_2^0$  and  the singlet  field  $S$ enter  the
$b$-squark     masses     only     through     the     mixing     term
$\left(A_b\Phi_1^0-\lambda S  \Phi_2^0\right)$.  As a  result of this,
it is easy to derive that $\partial m^2_{\tilde b_i}/\partial \phi_S =
\left(v_2/v_S\right)\;   \partial  m^2_{\tilde   b_i}/\partial  \phi_2
$. With the  aid of the latter, we  then find that $\Delta_b^{\phi_2}$
is   related   to    $\Delta_b^{\phi_S}$   by   $\Delta_b^{\phi_S}   =
\frac{v_2}{v_S}\;  \Delta_b^{\phi_2}$, if the  gluino-exchange diagram
is  only considered.  In  the same  approximation, a  similar relation
holds  true for the  pseudoscalar 1-loop  couplings:~$\Delta_b^{a_S} =
\frac{v_2}{v_S}\;  \Delta_b^{a_2}$.    However,  this  simple  scaling
behaviour  is   broken  by  the  chargino-exchange   diagram,  as  the
background Higgs  fields $\Phi^0_2$ and $S$ enter  the chargino masses
and  mixing angles  in a  non-linear manner  [cf.~(\ref{chmixing}) and
(\ref{mchi12})].

Finally, it is  worth commenting on the fact that  the coupling of the
neutral  would-be   Goldstone  boson  $G^0$   to  the  $b$   quark  is
proportional  to  the $b$-quark  mass,  as  a  consequence of  a  Ward
identity  involving   the  $Zb\bar{b}$-coupling.   Specifically,  the
coupling parameter $\Delta_b^{G^0}$, which is computed by
\begin{equation}
\Delta_b^{G^0}\ =\ i\sqrt{2}\, \left<{\partial\over\partial G^0}
\Delta_b\right>
\ =\ 
i\sqrt{2}\, \left<\left(c_\beta{\partial\over\partial
  a_1}\ +\ s_\beta{\partial\over\partial a_2}\right)\Delta_b\right>\,, 
\end{equation} 
is given by
\begin{equation}
\Delta_b^{G^0}\ =\ {\sqrt{2}\over v}\ \left<\Delta_b\right>\ .
\end{equation} 
Consequently, the $G^0b\bar{b}$-coupling has the tree-level SM form in
the limit of zero external momentum.

\subsection{Effective $\tau$-lepton Yukawa Couplings}
\label{tauyukawacouplings}

The derivation of effective  $\tau$-lepton Yukawa couplings goes along
the lines  discussed above  for the $b$-quark  case.  At the  one loop
order, there  is now only one  $t_\beta$-enhanced diagram contributing
to $\Delta_\tau$, which originates from the chargino-Higgsino-exchange
diagram of Fig.~\ref{domcoup}.  The effective functional $\Delta_\tau$
pertinent to the $\tau$-lepton self-energy is given by
\begin{equation}
  \label{Dtau}
\Delta_\tau  =  -{\alpha_w\over 4\pi}
\Bigg[\; 
m_{\tilde\chi_1} 
{\mathcal V}^\dag_{11}{\mathcal U}^\ast_{12}
\;  
B_0\left(0,m_{\tilde\chi_1},M_{\widetilde{L}}\right)
+\ m_{\tilde\chi_2}
{\mathcal V}^\dag_{12}{\mathcal U}^\ast_{22}
\;  
B_0\left(0,m_{\tilde\chi_2},M_{\widetilde{L}}\right)\;\Bigg]\; , 
\end{equation} 
where $M_{\widetilde{L}}$ is the  soft SUSY-breaking mass term for the
left-handed sleptons and the loop function $B^0(p^2,a,b)$ at $p^2=0$
takes on the simple form

\begin{equation}
B_0(0,a,b)\ =\ -\, \ln \left({ab\over Q^2}\right)\ +\ 1\ +\
{a^2 + b^2\over b^2-a^2}\ln\left({a^2\over b^2}\right)\; ,
\end{equation} 
where $Q^2$ is the renormalization scale, which is conveniently taken
to be $Q^2=m_{\tilde \chi_2}M_{\tilde L}$.

Notice that the non-holomorphic couplings both receive renormalization
scale-dependent contributions. This should not be suprising, as the
couplings also contain contributions from the mixing of $\phi_{2,S}$
 with the holomorphic $\phi_1$. We have checked that all 
scale-dependent terms vanish in the zero mixing limit of $\phi_{2,S}
 \to \phi_1$, corresponding to $v_1 \to 0$.

As  was done  above for  the  $b$-quark case,  the 1-loop  Higgs-boson
couplings to $\tau^+\tau^-$ may be computed from~(\ref{Dtau}) by means
of the HLET, where  the dominant $\phi_S\tau^+\tau^-$ diagram is shown
in  Fig.~\ref{domcoup}.   In  extensions  that  include  right  handed
(s)neutrinos, a second diagram mediated by Higgsino exchange must also
be  considered.    This  new   contribution  may  be   significant  if
(s)neutrinos  are  not too  heavy.   Such  further  extensions may  be
studied elsewhere.


\setcounter{equation}{0}
\section{Phenomenological Discussion}

In  this  section we  analyze  the  implications  of the  loop-induced
parameters   $\Delta_f^{\phi_S}$    and   $\Delta_f^{a_S}$   for   the
Higgs-boson  couplings  and   for  the  Higgs-boson  phenomenology  in
general. As was already mentioned,  the 1-loop coupling of the singlet
Higgs boson to the $b$ quark and the $\tau$ lepton becomes significant
at large values of $\tan\beta$.  In our analysis, we adopt a benchmark
scenario where the singlet Higgs-boson effects get enhanced. Unless is
stated otherwise,  the default values  of the SUSY parameters  for our
benchmark scenario are
\begin{equation}
  \label{Bench}
\begin{array}{lll}
\mu\ =\ \frac{1}{\sqrt{2}}\,\lambda\, v_S\ =\ 110~{\rm GeV},\qquad & 
t_\beta\ =\ 50, & \\
M_{\widetilde Q} \ =\ 300~{\rm GeV}, & M_{\widetilde{L}}\ =\ 90~GeV, & \\
M_{\tilde t}\ =\ 600~{\rm GeV}, &
M_{\tilde b} \ =\ 110~{\rm GeV},\qquad\quad & 
                              M_{\tilde{\tau}}\ =\  200~{\rm GeV},\\
A_t\ =\ 1~{\rm TeV}, & A_b\ =\ 1~{\rm TeV}, & A_\tau\ =\ 1~{\rm TeV}, \\
M_1\ =\ 400~{\rm GeV}, & M_2\ =\ 600~{\rm GeV}, & M_3\ =\ 
400~{\rm GeV}\; . 
\end{array}
\end{equation}
Notice  that  an important  constraint  on  the  choice of  the  above
parameters comes from the LEP data. This constraint is included in our
analysis.

\begin{figure}[t!]
\begin{center}
\includegraphics[scale=1.25]{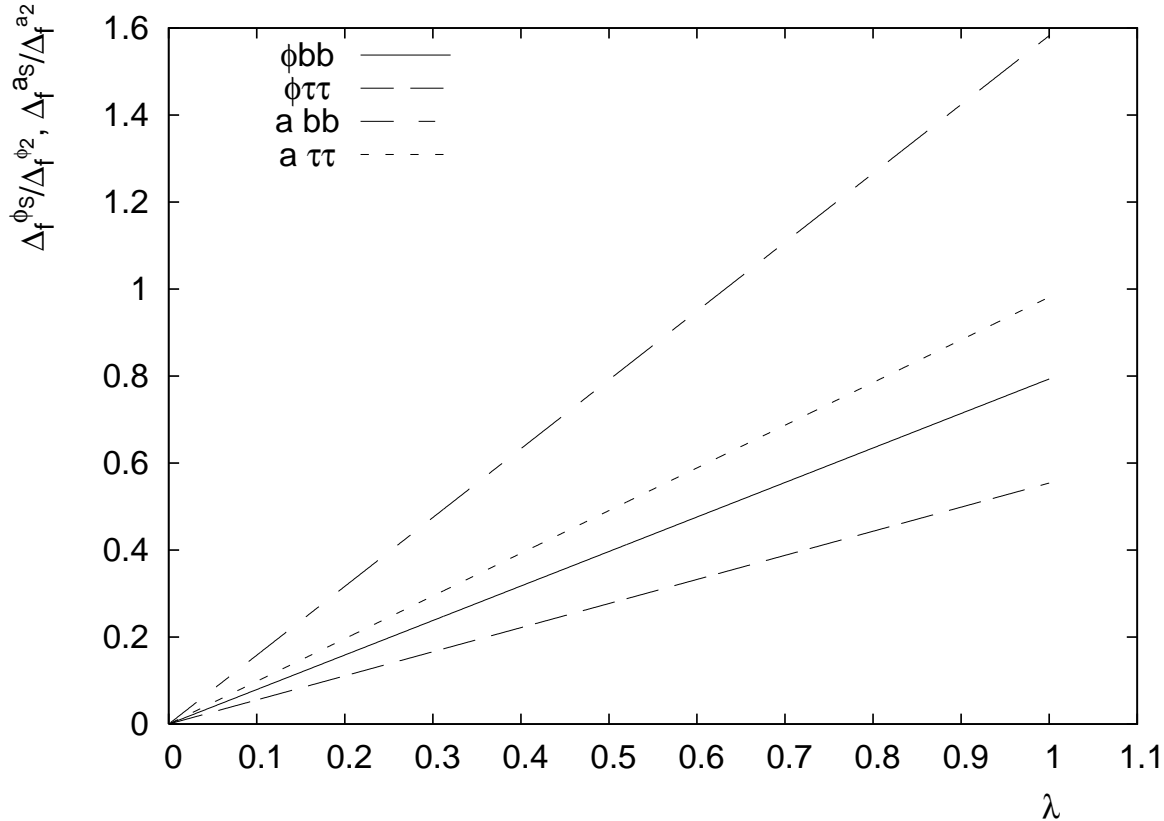}
\caption{\it      Numerical      estimates      of     the      ratios
  $\Delta^{\phi_S}_{b,\tau}/\Delta^{\phi_2}_{b,\tau}$               and
  $\Delta^{a_S}_{b,\tau}/\Delta^{a_2}_{b,\tau}$    as   functions   of
  $\lambda$. The values of the model parameters are given in~(\ref{Bench}).}
\label{ratios}
\end{center}
\end{figure}

Given  the   model  parameters~(\ref{Bench}),  the  coupling-parameter
ratios  $\Delta^{\phi_S}_f/\Delta^{\phi_2}_f$ and $\Delta^{a_S}_f/
\Delta^{a_2}_f$  are  shown  in Fig.~\ref{ratios},  as functions  of
 the  supersymmetric  coupling $\lambda$, keeping  the
$\mu$-parameter fixed.  Since the  radiative corrections to the Yukawa
couplings are dominated by SUSY  QCD effects in the case of the quarks, 
one might expect for the ratios  to be approximately given
by $\Delta^{\phi_S}_b/\Delta^{\phi_2}_b  \approx 
\Delta^{a_S}_b/\Delta^{a_2}_b  \approx v/v_S$. Specifically,
this ratio should reach the  value 1 for $\lambda\sim 0.65$.
As is illustrated in Fig.~\ref{ratios}, whilst this approximation is
valid for the psuedoscalar couplings, the subdominant corrections to 
the scalar couplings do not share this  simple scaling  behaviour and 
so  give rise to  a somewhat different relative magnitude for 
$\Delta^{\phi_S}_b/\Delta^{\phi_2}_b$.
Clearly, including interference effects between the contributing terms,
 the coupling parameter $\Delta^{\phi_S}_b$ becomes comparable 
with $\Delta^{\phi_2}_b$ only for large values of $\lambda$.

\vfill\eject

As  can  be  seen  from  the  effective  Lagrangian~(\ref{Leff}),  the
physical Higgs-boson couplings to the $b$ quark and the $\tau$ lepton,
e.g.~$H_{1,2,3}  f\bar{f}$  and  $A_{1,2}  f\bar{f}$, consist  of  two
contributions.    The  first   contribution  is   the   proper  vertex
interaction, which  is dominated by  the tree-level $\phi_1$-coupling.
The second contribution is the  mixing of the fields $\phi_{2,S}$ with
the $\phi_1$.  Such a mixing of  Higgs states occurs at the tree level
and is very significant for generic Higgs-boson mass matrices, as only
a   $2\%$    component   of   $\phi_1$   will    give   an   effective
$\phi_{2,S}$-coupling of order  $h_b^{\rm SM}$ at $t_\beta=50$.  Since
our interest is to assess the significance of the 1-loop singlet-Higgs
vertex effects, we will mainly focus  on variants of the mnSSM and the
NMSSM,  where  the  mixing  of  $\phi_1$ with  the  other  scalars  is
suppressed.

\vfill\eject

\subsection{Decoupling via a Heavy Charged Higgs Boson}

One way to switch  off the Higgs-boson self-energy transitions $\phi_1
\to \phi_{2,S}$ and/or $a_1 \to a_{2,S}$ is to consider the decoupling
of the states $\phi_1$ and $a_1$ via a heavy charged Higgs boson, with
$M_{H^\pm}   \stackrel{>}{{}_\sim}   M_{\rm   SUSY}$.   Although   the
decoupling  of $\phi_1$  may be  easily achieved  within the  MSSM for
relatively  large values  of $\tan\beta$~\cite{Mrenna},  the situation
becomes a bit more involved  in its singlet extensions. In particular,
for the CP-even Higgs sector, one has to arrange that both mass-matrix
elements   $\left(M_S^2\right)_{12}$   and   $\left(M_S^2\right)_{13}$
vanish,  which is  more difficult.   However, this  difficulty  is not
present  for the CP-odd  Higgs sector,  where the  mass-matrix element
$\left(M_P^2\right)_{12}$  could  vanish for  certain  choices of  the
model parameters,  thereby decoupling the CP-odd state  $a_1$ from the
rest of the mass spectrum.

\subsubsection{Mixing in the mnSSM}

Before  we discuss  the Higgs-mixing  effects, we  first give  a brief
overview of the Higgs sector of the mnSSM.  The renormalizable part of
the mnSSM superpotential is given by
\begin{eqnarray}
  \label{WmnSSM}
{\mathcal W}_{\rm mnSSM} &= & h_l\widehat H_d^T i\tau_2\widehat
L\widehat E\ +\ h_d\widehat H_d^T i\tau_2\widehat Q\widehat D\ +\ 
h_u\widehat Q^T i\tau_2\widehat H_u\widehat U\nonumber\\
&&+\; \lambda\widehat S\widehat H_d^T i\tau_2\widehat H_u\ +\ 
t_F \widehat S\; .
\end{eqnarray} 
In~(\ref{WmnSSM})  the  term linear  in  $\widehat{S}$  is induced  by
supergravity quantum effects from Planck-suppressed non-renormalizable
operators in the K\"ahler  potential and superpotential.  Depending on
the  discrete $R$  symmetries  imposed on  the  theory, the  effective
tadpole  parameter $t_F$  and  its associate  soft SUSY-breaking  term
$t_S\, S$ may  be generated at loop levels higher than  5 and can both
be of order  $M_{\rm SUSY}$.  These two interactions  are essential to
break  the  unwanted PQ  symmetry.   Further  details  related to  the
tadpole generation  and the breaking of  the PQ symmetry  may be found
in~\cite{comment}.

In the mnSSM, the tree-level CP-odd mass matrix $M_P^2$ reads
\begin{eqnarray}
  \label{MPmnSSM}
\left(M_P^2\right)_{11} & = & M_a^2\;, \nonumber\\
\left(M_P^2\right)_{12} & = & 
{v\over v_S} \left( s_\beta c_\beta M_a^{2}\: +\: 
m_{12}^2 \right)\; ,\nonumber\\ 
\left(M_P^2\right)_{22} & = & {v^2\over v^2_S}\left(s_\beta
c_\beta M_a^2\: +\: m_{12}^2\right)\ +\ {\lambda t_S\over\mu}\ , 
\end{eqnarray}
where  subscript 1  refers to  the CP-odd  state $a  = -s_\beta  a_1 +
c_\beta a_2$ and the subscript 2 to the CP-odd state $a_S$.  Moreover,
$M_a$ is  the would-be  MSSM pseudoscalar Higgs  mass, related  to the
charged   Higgs-boson  mass   by  $M_a^{2}=M_{H^\pm}^{2}-M_W^2+{1\over
2}\lambda^2 v^2$ at  the tree level, and $m_{12}^2$  is related to the
effective superpotential tadpole  $t_F$ by $m_{12}^2=\lambda t_F$.  It
is  important to  comment  that  the dominant  scalar  top and  bottom
corrections to the  CP-odd mass matrix $M^2_P$ can  all be absorbed to
$M_a$ and so do not modify its tree-level form.

\begin{figure}[t!]
\begin{center}
\includegraphics[scale=1.25]{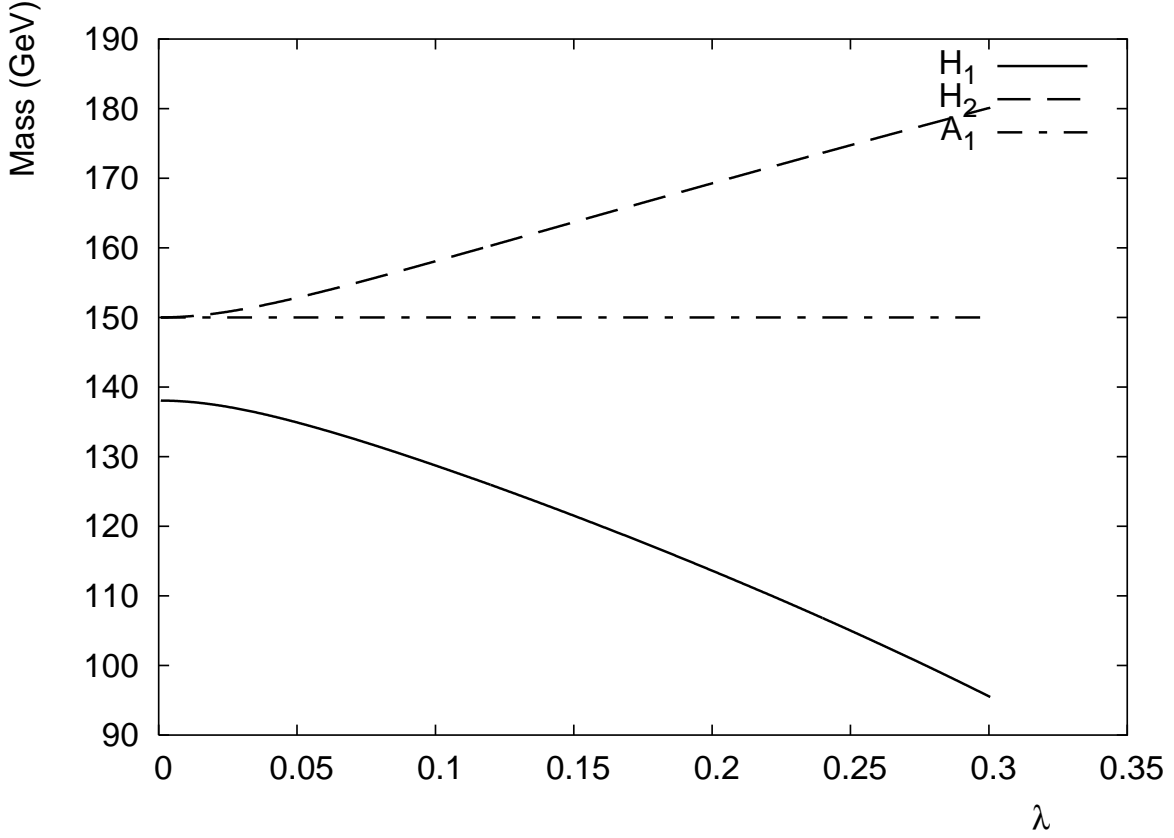}
\caption{\it Masses of the $H_1$~(solid line), $H_2$~(dashed line) and
$A_1$~(dot-dashed line) bosons in the mnSSM with $\mu=110$ GeV,
$\lambda t_S/\mu=150^2$ GeV$^2$ and $m_{12}^2=-0.5$ TeV$^2$. 
The values of other soft SUSY-breaking
parameters are given in~(\ref{Bench}).}
\label{masstwo}
\end{center}
\end{figure}

By analogy, the tree-level CP-even mass matrix $M^2_S$ is given by
\begin{eqnarray}
\left(M^2_S\right)_{11} &=& c_\beta^2 M_Z^2\: +\: s_\beta^2 M_a^2\; ,
     \nonumber\\
\left(M^2_S\right)_{12} & = & -s_\beta c_\beta\left(M_a^2 +M_Z^2 -
\lambda^2 v^2\right)\; ,\nonumber\\
\left(M^2_S\right)_{13} & = & -{v\over v_S}\left(s_\beta^2 c_\beta
M_a^2 + s_\beta m_{12}^2 - 2c_\beta\mu^2\right)\; ,\nonumber\\
\left(M^2_S\right)_{22} & = & s_\beta^2 M_Z^2 + c_\beta^2 M_a^2\; , \nonumber\\
\left(M^2_S\right)_{23} & = & -{v\over v_S}\left(s_\beta c_\beta^2
M_a^2 + c_\beta m_{12}^2 - 2s_\beta\mu^2\right)\; ,\nonumber\\
\left(M^2_S\right)_{33} & = & s_\beta c_\beta\left({v\over
  v_S}\right)^2 \left(s_\beta c_\beta M_a^2 + 
m_{12}^2\right)+{\lambda t_S\over\mu}\ ,
\end{eqnarray}
where   the   subscripts  $1,2,3$   refer   to   the  CP-even   states
$\phi_{1,2,S}$,  respectively.   In  our  numerical analysis  we  also
include  the 1-loop  corrections to  $M^2_S$  due to  both (s)top  and
(s)bottom loops,  which play an  important role both  for intermediate
and large values of $t_\beta$.

\pagebreak

To  cancel the  mixing  of state  $a_1$  with the  other CP-odd  Higgs
states, we only need to  choose $m_{12}^2$, such that it satisfies the
relation
\begin{equation}
s_\beta c_\beta M_a^2\ +\ m_{12}^2\ =\ 0\; . 
\end{equation}
This relation also  approximately cancels $\left(M_S^2\right)_{13}$ in
the large  $t_\beta$ limit.  We  shall enforce this constraint  at all
times when considering the mnSSM.

\begin{figure}[t!]
\begin{center}
\includegraphics[scale=1.25]{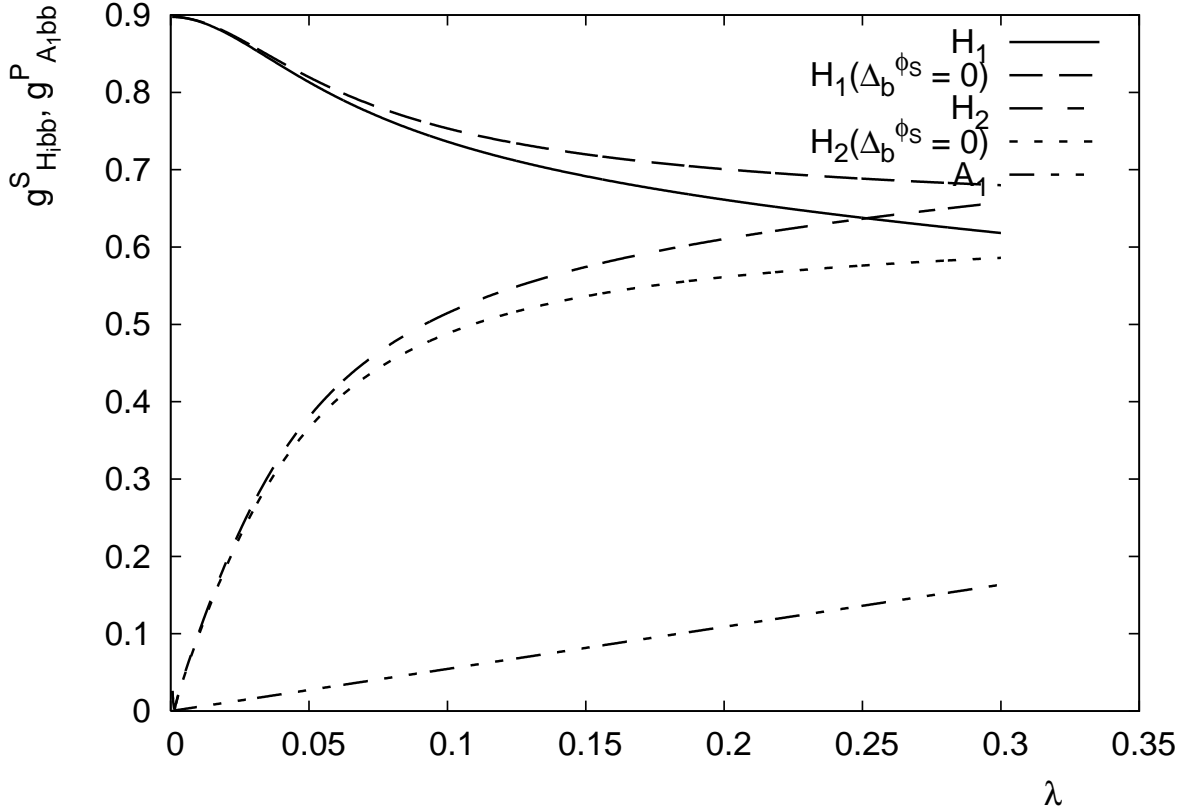}
\caption{\it  The SM-normalized  couplings $H_1b\bar  b$,
$H_2b\bar b$ and $A_1b\bar b$ in the
mnSSM, as  functions of  $\lambda$, for the  same model  parameters as
in~Fig.~\ref{masstwo}. Also shown are the corresponding couplings 
$H_1b\bar b$ and $H_2b\bar b$ with the singlet threshold correction
$\Delta^{\phi_S}_b=0$ for comparison. The pseudoscalar couplings
are approximately zero when $\Delta^{a_S}_b=0$.}
\label{couplingtwo}
\end{center}
\end{figure}

In  Fig.~\ref{masstwo}  we display  the  masses  of  the two  lightest
CP-even Higgs  bosons $H_1$  and $H_2$ and  the lightest  CP-odd Higgs
boson  $A_1$.  Since  we  have taken  $M_{H^\pm}=5$  TeV and  $\lambda
t_S/\mu=150^2$   GeV$^2$,  the   remaining   physical  Higgs   states,
$H_3\sim\phi_1$ and  $A_2\sim a$, are heavy of  order $M_{H^\pm}$.  We
observe that the  lightest Higgs boson mass $M_{H_1}$  goes well below
the LEP limit from direct Higgs searches, for large values of $\lambda
\stackrel{>}{{}_\sim} 0.3$.   Hence, the coupling $\lambda$  has to be
smaller  than  about 0.3  for  a  phenomenological  viable model.   In
Fig.~\ref{couplingtwo} we  then show  the dependence of  the $b$-quark
Yukawa   couplings  $g^S_{H_{1,2}bb}$   and  $g^P_{A_1bb}$,   for  the
aforementioned  scenario. We  find that  the CP-even  Yukawa couplings
$g^S_{H_{1,2}bb}$   receive   appreciable   contributions   from   the
tree-level mixing  of the state  $\phi_1$ with $\phi_{2,S}$,  which is
competitive      to     the     loop-induced      Yukawa     couplings
$\Delta^{\phi_{2,S}}_b$.   On the other  hand, the  coupling $g^P_{A_1
b\bar  b}\approx g^P_{a_S  b\bar b}$  is completely  dominated  by the
1-loop   contribution  $\Delta^{a_S}_b$.    For  moderate   values  of
$\lambda\sim 0.3$, we  find that $g^P_{A_1 b\bar b}  \sim 0.15$, so the
$A_1\bar{b}b$-coupling is  $\sim 15\%$ of the SM  Higgs boson coupling
to the $b$  quark. Moreover, the decay $A_1  \to \bar{b}b$ is expected
to  be the dominant  decay channel  in this  specific scenario  of the
mnSSM.

\begin{figure}[t!]
\begin{center}
\includegraphics[scale=1.25]{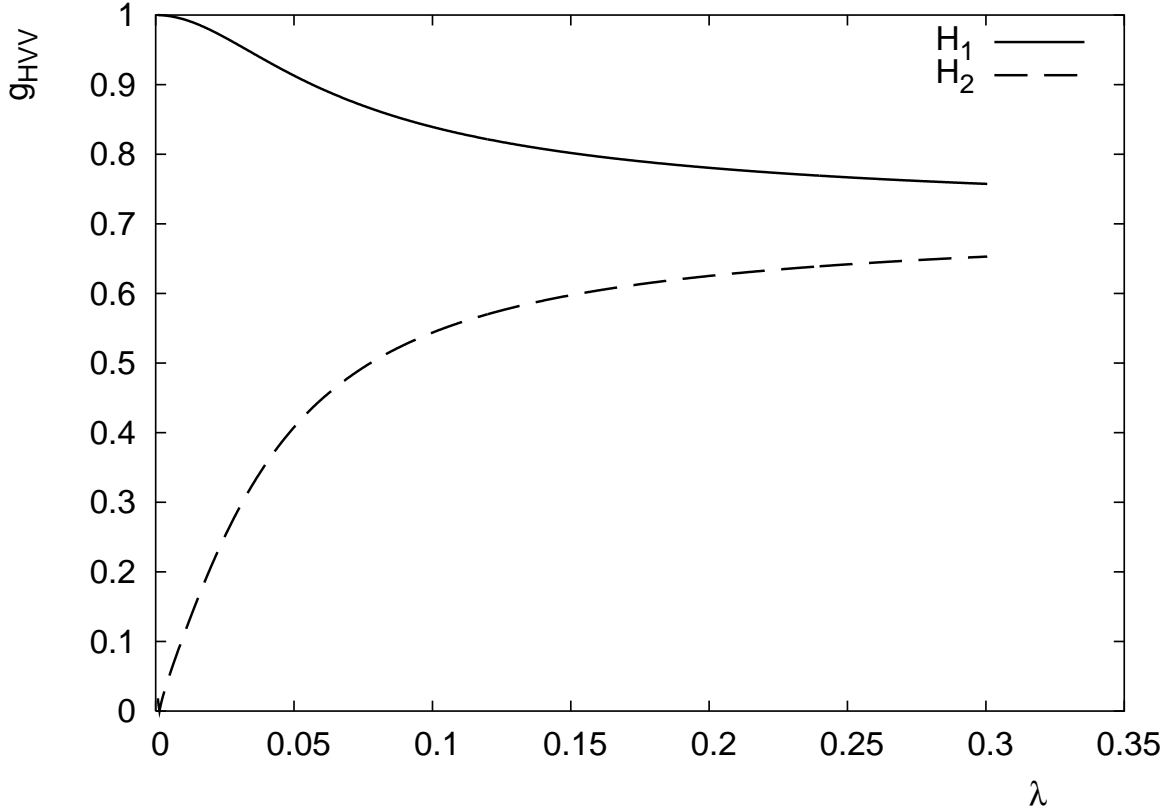}
\caption{\it  Effective vector  boson couplings  for the  Higgs bosons
$H_1$~(solid line) and $H_2$~(dashed line)  in the mnSSM, for the same
model parameters as in~Fig~\ref{masstwo}.}
\label{ghvvtwo}
\end{center}
\end{figure}

For  completeness, we  present numerical  estimates for  the effective
Higgs-boson  couplings to  the $W^\pm$  and $Z$  bosons. These  can be
determined by the effective Lagrangian
\begin{equation}
  \label{LHVV}
{\mathcal  L}_{HVV}\ =\ g_w  M_W\sum_{i=1}^3 g_{H_iVV}\left(H_iW^+_\mu
W^{-,\mu}\ +\ {1\over c_w^2}H_iZ_\mu Z^\mu\right)\; ,
\end{equation} 
where $c_w=\sqrt{1-s_w^2}=M_W/M_Z$ and
\begin{equation}
g_{H_iVV}\ =\ O^H_{2i}\, c_\beta\: +\: O^H_{1i}\, s_\beta\; .
\end{equation}
In  the  scenario of  the  mnSSM  specified  above, the  SM-normalized
couplings $g_{H_{1,2}VV}$  are shown in  Fig.~\ref{ghvvtwo}. Combining
the   results   of   this   last   figure   with   Figs.~\ref{masstwo}
and~\ref{couplingtwo}, we observe that both the lightest CP-even Higgs
bosons $H_{1,2}$  will predominantly decay into $b$  quarks, for small
values of $\lambda$, e.g.~for $\lambda \sim 0.05$.  For larger values,
i.e.~for  $\lambda \stackrel{>}{{}_\sim}  0.1$, only  the  $H_1$ boson
will decay into  $b$ quarks, whereas the heavier  one $H_2$ will decay
into $W^\pm$ bosons.


\subsubsection{Mixing in the NMSSM}

We now  turn our  attention to the  NMSSM. The superpotential  of this
model is given by
\begin{eqnarray}
{\mathcal W}_{\rm NMSSM} &=&
h_l \widehat H_d^T i\tau_2 \widehat L\widehat E\: +\: 
h_d \widehat H_d^T i\tau_2\widehat Q\widehat D\: +\: 
h_u \widehat Q^T i\tau_2\widehat H_u\widehat U\nonumber\\
&&+\: \lambda \widehat S\widehat H_d^T i\tau_2\widehat H_u\: 
+\: {\kappa\over 3}\widehat S^3\; .
\end{eqnarray} 
The difference between  the mnSSM and the NMSSM  is that the effective
tadpole parameter linear  in $\widehat{S}$ in the former  model is now
replaced by  an operator cubic  in $\widehat{S}$.  In addition  to the
superpotential  term ${\kappa\over  3}\widehat S^3$,  there will  be a
soft  SUSY-breaking operator  ${\kappa\over  3}\,A_\kappa S^3$,  which
needs be considered as well in the calculation of the Higgs-boson mass
matrices.

In the same  weak basis as the one considered for  the mnSSM, we first
present  the CP-odd  Higgs-boson  mass matrix  $M^2_P$.   At the  tree
level, $M^2_P$ may be conveniently expressed as
\begin{eqnarray}
  \label{MPnmssm}
\left(M_P^2\right)_{11} & = & M_a^2\;, \nonumber\\
\left(M_P^2\right)_{12} & = & {v\over v_S}\left(s_\beta c_\beta M_a^2
\: +\: 3{\kappa\over\lambda}\mu^2\right)\; ,\nonumber\\
\left(M_P^2\right)_{22} & = & 
{v^2\over v_S^2}s_\beta c_\beta \left( s_\beta c_\beta M_a^2\: -\:
3{\kappa\over\lambda}\mu^2\right)\ +\ 3{\kappa\over\lambda}\mu
A_\kappa\; .
\end{eqnarray} 
Here $s_\beta c_\beta M_a^2=\left(\mu A_\lambda - 
{\kappa\over\lambda}\mu^2\right)$ at tree level.
In similar  fashion, the individual matrix elements  of the tree-level
CP-even Higgs-boson mass matrix $M^2_S$ are written down as follows:
\begin{eqnarray}
\left(M_S^2\right)_{11} &=& c_\beta^2 M_Z^2\: +\: 
                                              s_\beta^2 M_a^2\; ,\nonumber\\
\left(M_S^2\right)_{12} &=& -\,s_\beta c_\beta\left( M_a^2\: +\: M_Z^2\: 
-\: \lambda^2 v^2\right)\; ,\nonumber\\
\left(M_S^2\right)_{22} &=& s_\beta^2 M_Z^2\: +\: 
                                              c_\beta^2 M_a^2\; ,\nonumber\\
\left(M_S^2\right)_{13} &=& -\, {v\over v_S}\, 
\left( s_\beta^2 c_\beta M_a^2\: -\: 2 c_\beta\mu^2\: -\; 
{\kappa\over\lambda}\, s_\beta\mu^2\, \right)\; ,\nonumber\\
\left(M_S^2\right)_{23} &=& 
-{v\over v_S}\left(s_\beta c^2_\beta M_a^2\: -\: 2 s_\beta\mu^2\: -\: 
{\kappa\over\lambda}c_\beta\mu^2\right)\; ,\nonumber\\
\left(M_S^2\right)_{33} &=& {v^2\over v_S^2}\, 
s_\beta c_\beta\, \left(\, s_\beta c_\beta M_a^2\: +\: 
{\kappa\over\lambda}\,\mu^2\right)\: -\: 
{\kappa\over\lambda}\mu A_\kappa\: +\: 4{\kappa^2\over\lambda^2}\mu^2\, .
\end{eqnarray}


\begin{figure}[t!]
\begin{center}
\includegraphics[scale=1.25]{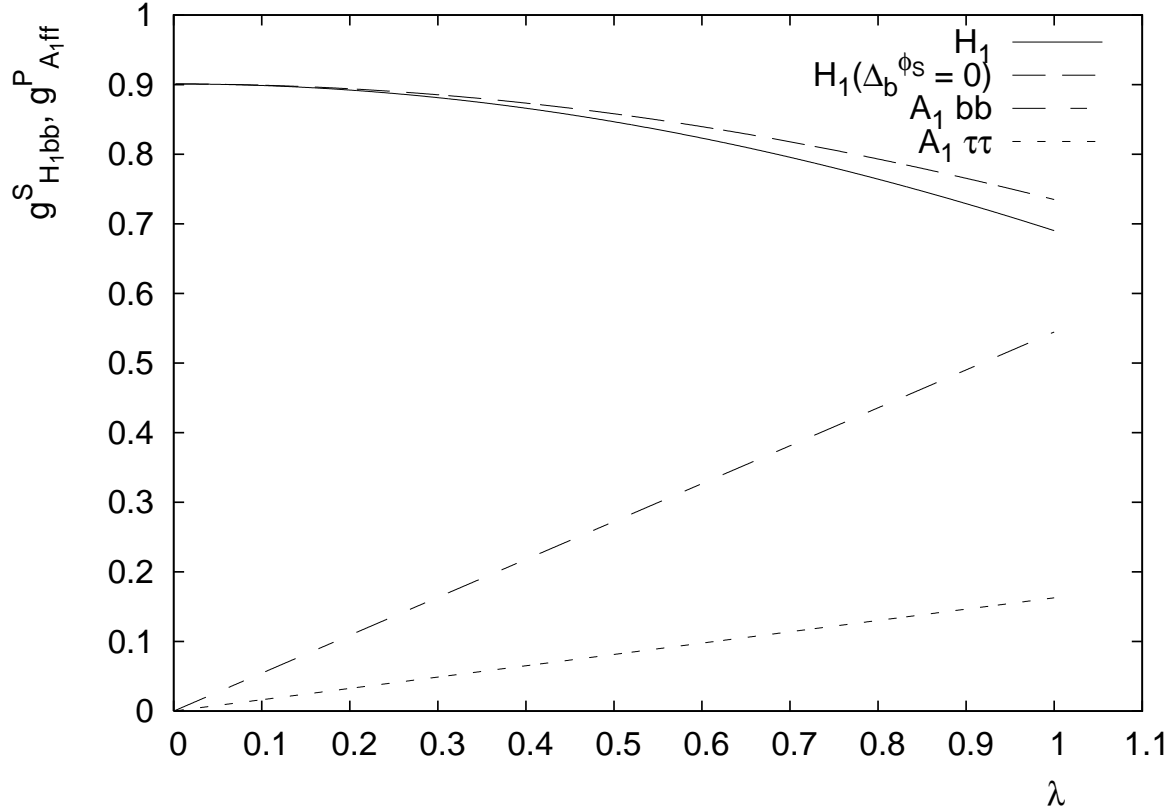}
\caption{\it The SM-normalized couplings $H_1b\bar b$~(solid
  line), $A_1b\bar b$~(dashed line)  and
  $A_1\tau^+\tau^-$~(dot-dashed line) in the NMSSM, as functions of
$\lambda$, 
   where the parameters $\kappa$ and $A_\kappa$ are constrained 
by~(\ref{kl}) and~(\ref{Ak}). We also show the coupling 
$H_1b\bar b$ with the singlet threshold correction
$\Delta^{\phi_s}_b=0$. Both pseudoscalar couplings $A_1ff$ are 
approximately zero when $\Delta^{a_s}_f=0$.}
\label{couplingone}
\end{center}
\end{figure}


Following the same  rationale as in the mnSSM,  we suppress the mixing
of  the CP-odd state  $a \approx  -a_1$ with  $a_S$ by  requiring that
$\left( M^2_P \right)_{12} =  0$. Hence, we find that
\begin{equation}
  \label{kl}
{\kappa\over\lambda}\ =\ -\,{s_\beta c_\beta \over 3\mu^2}\, M_a^2\; .
\end{equation} 
Note that  for positive $\lambda$,  we must have negative  $\kappa$ to
fulfil  the  above   constraint.   Substituting~(\ref{kl})  into  the
expression for  $\left(M_P^2\right)_{22}$ given in~(\ref{MPnmssm}), we
get the mass for the lightest CP-odd Higgs boson $A_1$, i.e.
\begin{equation}
M^2_{A_1}\ =\ \left(M_P^2\right)_{22}\ =\
-\, 3\, {\kappa\over\lambda}\; \mu\, \left[2\left(v\over
  v_S\right)^2s_\beta c_\beta\mu\: -\: A_\kappa\, \right]\; .
\end{equation}
For  our  benchmark  value  of  $\mu  =  110$  GeV  and  for  moderate
$\lambda\sim  0.6$  with  $t_\beta=50$,  the positivity  condition  on
$M^2_{A_1}$ gives an upper bound on $A_\kappa\sim 5$ GeV. The simplest
option would be to set  $A_\kappa=0$.  The mass of $A_1$ then strongly
depends upon the  value of $\lambda$, through the  factor $v/v_S$.  In
order to examine the effect of the threshold corrections on very light
singlets  across  a  larger  range  of  $\lambda$,  we  instead  allow
$A_\kappa$ to vary such that
\begin{equation}
 \label{Ak}
2\left(v\over v_S\right)^2s_\beta c_\beta\mu\: -\: A_\kappa\
=\ 0.05\sim 0.06
\end{equation} 
This, together  with the above constraint~(\ref{kl}),  gives a singlet
pseudoscalar mass  of $M_{A_1} =  6\sim 9$ GeV for  $M_{H^\pm}=2$ TeV,
across the full  range of $\lambda$. The couplings  of $H_1$ to $b\bar
b$ and of  $A_1$ to both $b\bar b$ and  $\tau^+\tau^-$ pairs are shown
in  Fig.~\ref{couplingone}. In  this  variant of  the NMSSM,  lightest
CP-even   Higgs-boson    mass   $M_{H_1}$   is    within   the   range
$120$--$140$~GeV across the full range of~$\lambda$.

The condition~(\ref{kl}) can be re-written as a constraint on the
parameter $A_\lambda$,

\begin{equation}
A_\lambda=-2{\kappa\over\lambda}\mu={2s_\beta c_\beta\over 3\mu}
M^2_a
\label{A_lambda}
\end{equation}
In view of (\ref{A_lambda}) it is clear that a singlet 
dominated light pseudoscalar is present in the NMSSM
spectrum if $A_\lambda\sim 200$~GeV and $A_\kappa\sim 5$~GeV. This can
be naturally arranged in gauge or gaugino mediated SUSY breaking
scenarios, where $A_\lambda$ and $A_\kappa$ are zero at tree level.
Quantum contributions from gaugino masses produce non-zero values at 
the one- and two-loop level respectively, leading to the approximate
scales shown above if the gaugino masses are of the order $100$~GeV
\cite{Dermisek:2005ar}. Although we have artificially set
$(M_P^2)_{12}$ exactly to zero in order to make the effect of the
one-loop correction explicit, similar considerations apply to the
more natural scenario of small but non-vanishing pseudoscalar mixing.

In the last  years, there has been some  interest in the phenomenology
of  light Higgs pseudoscalars,  which may  provide an  invisible decay
channel for a light SM-like Higgs boson.  If these CP-odd scalars have
a  large  singlet  component,  it  is  possible  for  them  to  escape
experimental  bounds  \cite{GHE}.   It  is clear  that  the  threshold
corrections can have a significant effect on the branching ratios of a
light CP-odd singlet scalar for moderate to large values of $\lambda$.
Previous studies have considered  detection of these particles through
decays to photon pairs  as the dominant mode \cite{Dobrescu:2000jt} in
the limit  of vanishing singlet-doublet  pseudoscalar mixing. However,
our analysis shows  that this need not be the case,  and the impact of
the hadronic decays of $A_1$ in so-called ``invisible Higgs" scenarios
should still be considered even in this limit.


\subsection{Decoupling by Tuning $\lambda$ in the mnSSM}

In  the mnSSM the  turning off  of the  singlet-doublet mixing  of the
CP-odd Higgs scalars leads automatically  to a suppression of the mass
matrix element $(M^2_S)_{13}$, which in turn implies a small mixing of
$\phi_1$ with $\phi_S$. Nevertheless,  it is also possible to decouple
$\phi_1$  from  the  other  CP-even  Higgs state  $\phi_2$  by  tuning
$\lambda$. Hence one may choose a value for $\lambda$, such that it is
$(M^2_S)_{12} = 0$, or equivalently
\begin{equation}
-s_\beta c_\beta\left(M_a^2\: +\: M_Z^2\: -\: 
\lambda^2v^2\right)\: +\: \delta_{\rm rad}\ =\  0\;,
\end{equation} 
where $\delta_{\rm rad}$ represents  radiative corrections to the tree
level  mass  matrix. Evidently,  in  the  absence  of any  fundamental
reason, such a scenario should be considered to be somewhat contrived,
as it  relies upon an unnatural  cancellation of different  terms to a
relatively high level of precision.

\begin{figure}[t!]
\begin{center}
\includegraphics[scale=1.25]{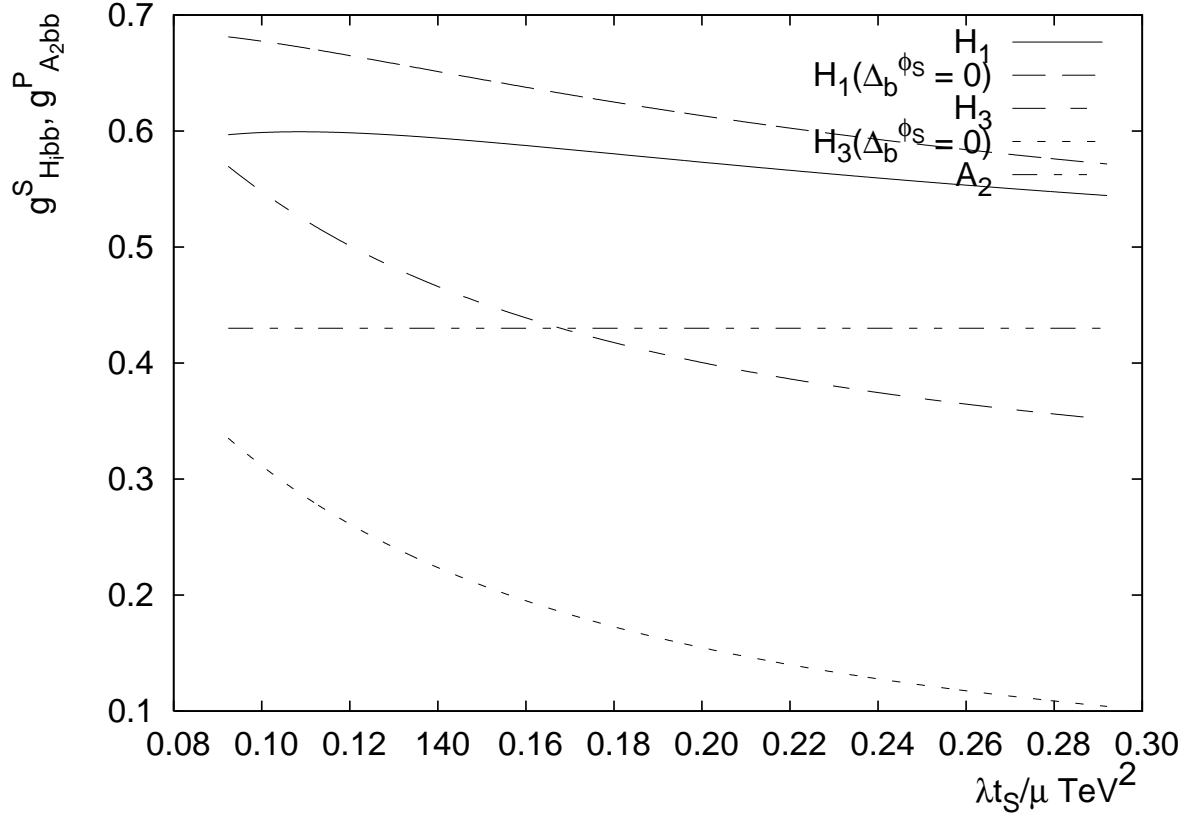}
\caption{\it  Effective  $b$-quark Yukawa couplings  of  the  Higgs  
scalars $H_1$,  $H_3$ and  $A_2$, as functions of $\lambda t_S/\mu$
, where $\mu=110$ GeV, $M_{H^\pm}=120$ GeV, $m_{12}^2=-292$ GeV$^2$
 and $\lambda=0.7895$. Also shown are the corresponding couplings 
$H_1b\bar b$ and $H_3b\bar b$ with the singlet threshold correction
 $\Delta_b^{\phi_S}=0$ for comparison. The pseudoscalar couplings
are approximately zero when $\Delta_b^{a_S}=0$.}
\label{novellimit}
\end{center}
\end{figure}

In Fig.~\ref{novellimit} we present  numerical estimates of the Yukawa
couplings  for  this  contrived  model.   As  input  values,  we  take
$M_{H^\pm}=120$  GeV  and   $\lambda=0.7895$.   The  $b$-quark  Yukawa
couplings  of $H_1\sim\phi_2$, $H_3\sim\phi_S$  and $A_2\sim  a_S$ are
plotted as functions of singlet mass parameter $\lambda t_S/\mu$.  The
threshold  corrections  are  independent  of the  singlet  mass  scale
$\lambda t_S/\mu$, so we expect a flat line when these dominate, as is
the case  for the  CP-odd Higgs field  $A_2$. Instead,  the respective
couplings of the CP-even Higgs bosons show a noticeable variation with
increasing $\lambda  t_S/\mu$, which originate from the  fact that the
mass matrix elements $(M^2_S)_{13}$  and $(M^2_S)_{23}$ do not exactly
cancel in this contrived model.  As a consequence, the Yukawa coupling
constant   $g_{H_1\bar  bb}\sim  g_{\phi_2   \bar  bb}$   receives  an
appreciable contribution thanks to the mixing between the doublets via
an intermediate  singlet state,  i.e.~$\phi_2 \to \phi_S  \to \phi_1$.
Considering LEP constraints, this scenario of the mnSSM is only viable
for heavy  Higgs singlets with  masses around 300~GeV. 

\bigskip
\setcounter{equation}{0}
\section{Conclusions and Future Directions}

Minimal  extensions  of the  MSSM  generically  include singlet  Higgs
bosons.   Although  singlet Higgs  bosons  have  no  direct or  proper
couplings  to the SM  particles, their  interaction with  the observed
matter can still be significant as a result of two contributions.  The
first one is their mixing with  Higgs doublet states, which is the one
often considered in the  literature.  The second contribution is novel
and persists  even if the  Higgs doublet-singlet mixing  is completely
switched  off.  It results  from gluino,  chargino and  squark quantum
effects at the 1-loop level and has been the focus of this paper.

In  this  article  we  have  derived  an  effective  Lagrangian  which
describes the interactions between  the Higgs bosons and the down-type
quarks and  leptons in CP-conserving  singlet extensions of  the MSSM.
We have found that  the loop-induced singlet Higgs-boson couplings are
enhanced for large values of $\tan\beta \stackrel{>}{{}_\sim} 40$.  We
have examined the  effects of these couplings on  the phenomenology of
two  such models,  the mnSSM  and the  NMSSM. Although  the  mixing of
$\phi_1$  with the  other CP-even  Higgs scalars  $\phi_{2,S}$  is the
leading effect for most of the parameter space, we have found that the
threshold corrections to the  Yukawa couplings remain relevant and can
play an important role in  the phenomenology of possibly light singlet
Higgs scalars.

In the absence  of a Higgs doublet-singlet mixing,  the 1-loop quantum
effects we have been studying here will be the only means by which the
CP-odd singlet  may couple to  quarks and leptons. For  a sufficiently
light CP-odd singlet  scalar, with a mass below  the squark threshold,
the  loop-induced Yukawa  couplings  will provide  its dominant  decay
channel  into   $b$  quarks.   This   has  important  phenomenological
implications,  since  a  SM-like  Higgs  boson will  no  longer  decay
invisibly into  a pair of  light singlet pseudoscalars  $A_1A_1$.  The
threshold corrections  that we calculated  here will give rise  to the
leading decay  mode $A_1  \to b\bar{b}$. In  fact, this  decay channel
depends on the  mass of $A_1$, and can have a  coupling strength of up
to $\sim 1/2$ of the corresponding SM coupling, i.e.~$g^P_{A_1 bb}\sim
0.5$.

There  are several  possible  new directions  for  future study.   For
instance,  one may  lift  the  assumption of  CP  conservation in  the
singlet  extensions  of the  MSSM.   Then,  light  CP-odd scalars  may
contribute  to electron  and neutron  electric dipole  moments  at the
2-loop level~\cite{CKP}.  It would  be interesting to study the impact
of  those  contributions in  the  presence  of CP-violating  threshold
corrections.   Another possible  direction  for future  investigations
will be to  calculate the off-diagonal couplings of  the singlet Higgs
bosons  to down-type  quarks~\cite{Hiller}.  Our  effective Lagrangian
presented here  may be  generalized to include  these flavour-changing
neutral-current (FCNC) interactions of the Higgs bosons to quarks.  It
would be  particularly valuable to  explore the impact of  the singlet
Higgs-boson FCNC effects on $K$- and $B$-meson observables.

\subsection*{Acknowledgements}

This  work  is  supported  in   part  by  the  PPARC  research  grant:
PP/D000157/1.

\newpage

\def\theequation{\Alph{section}.\arabic{equation}}

\begin{appendix}
\setcounter{equation}{0}
\section{Background-Field-Dependent Masses and Mixing Angles}

In this appendix we present our notation and conventions for the
masses and mixing angles which enter the calculations of
Sections~\ref{byukawacouplings} and~\ref{tauyukawacouplings}.
We have divided the appendix into two subsections. The first gives the
relevant squark mass parameters, whilst the second contains the 
corresponding information for the chargino sector.

\subsection{Squark Masses and Mixing Angles}

Neglecting D-term contributions, the scalar top and bottom mass matrices
may be written in the $(\tilde q_L, \tilde q_R)^T$ basis as

\begin{equation}
\widetilde{\mathcal M}^2_t = \left(
\begin{array} {cc}
\tilde M_Q^2 + h_t^2|\Phi_2^0|^2 
& h_tA_t\Phi_2^{0\ast}-h_t\lambda S \Phi_1^{0\ast}\\
h_tA_t\Phi_2^0-h_t\lambda S^\ast \Phi_1^0
& \tilde M_t^2 + h_t^2|\Phi_2^0|^2\end{array}\right)\; ,
\end{equation}

\begin{equation}
\widetilde{\mathcal M}^2_b = \left(
\begin{array} {cc}
\tilde M_Q^2 + h_b^2|\Phi_1^0|^2
& h_bA_b\Phi_1^0-h_b\lambda S \Phi_2^0\\
h_bA_b\Phi_1^{0\ast}-h_b\lambda S^\ast \Phi_2^{0\ast}
& \tilde M_b^2 + h_b^2|\Phi_1^0|^2\end{array}\right)\; .
\end{equation}
Where the fields $\Phi_i^0$ and $S$ have been defined after
(\ref{lagrangianone}).
These matrices are diagonalized by unitary matrices which may be
 parameterised as

\begin{equation}
\left(\begin{array}{c}
\tilde q_L\\
\tilde q_R\end{array}\right) = 
\left(\begin{array}{cc}
\cos\theta_q & \sin\theta_q\\
-\sin\theta_q & \cos\theta_q\end{array}\right)
\left(\begin{array}{cc}
1 & 0 \\
0 & e^{i\delta_q}\end{array}\right)
\left(\begin{array}{c}
\tilde q_1\\
\tilde q_2\end{array}\right)\; ,
\end{equation}
where 
\begin{eqnarray}
\delta_q & = & 
\arg\left(\left(\widetilde{\mathcal M}_q^2\right)_{12}
\right)\; ,\nonumber\\
s_t \equiv \sin\theta_t & = & 
\sqrt{m^2_{\tilde t_2}-\tilde M_Q^2-h_t^2|\Phi_2^0|^2\over
m^2_{\tilde t_2}-m^2_{\tilde t_1}}\; ,\nonumber\\
s_b \equiv \sin\theta_b & = &
\sqrt{m^2_{\tilde b_2}-\tilde M_Q^2-h_b^2|\Phi_1^0|^2\over
m^2_{\tilde b_2}-m^2_{\tilde b_1}}\; ,
\end{eqnarray}
and $c_q \equiv \cos\theta_q = \sqrt{1-s_q^2}$. 

The field-dependent scalar quark masses are given by

\begin{eqnarray}
\tilde m^2_{t_{1(2)}} & = &
{1\over 2}\left[\tilde M_Q^2+\tilde M_t^2+2h_t^2|\Phi_2^0|^2
+(-)\sqrt{\left(\tilde M_Q^2-\tilde M_t^2\right)^2
+4h_t^2|A_t\Phi_2^0-\lambda S^\ast\Phi_1^0|^2}\right]\; ,\\
\tilde m^2_{b_{1(2)}} & = &
{1\over 2}\left[\tilde M_Q^2+\tilde M_b^2+2h_b^2|\Phi_1^0|^2
+(-)\sqrt{\left(\tilde M_Q^2-\tilde M_b^2\right)^2
+4h_b^2|A_b\Phi_1^0-\lambda S\Phi_2^0|^2}\right]\; .
\end{eqnarray}

\subsection{Chargino Masses and Mixing Angles}
The Lagrangian  describing the chargino masses and their Yukawa 
interactions to Higgs bosons is given by
\begin{equation}
-\,\mathcal{L}_\chi^\pm \ =\
\left(-i\tilde w_{L}^-, \tilde h_{1L}^-\right)\; M\; 
\left(\begin{array}{c} -i\tilde w_{L}^+ \\
\tilde h_{2L}^+ \end{array}\right)\: +\: {\rm h.c.},
\end{equation} 
with
\begin{equation}
M\ =\ \left(\begin{array}{cc}
M_2 & g_w\Phi^{0*}_2 \\
g_w\Phi^0_1 & \lambda S \end{array}\right)\; .
\end{equation}
The mass matrix $M$ may be diagonalized by the bi-unitary transformation
\begin{equation}
M_D\ =\ {\mathcal U}^\ast\, M\, {\mathcal V}^\dag\; ,
\end{equation}
where $M_D = {\rm diag}( m_{\chi_1} , m_{\chi_2} )$. Both the chargino
masses and  the elements  of the unitary  matrices ${\mathcal U}$
 and  ${\mathcal V}$ depend
explicitly on the Higgs background fields $\Phi^0_{1,2}$ and $S$.

We may parameterise the unitary matrices ${\mathcal U},\; {\mathcal V}$
as
\begin{eqnarray}
{\mathcal U}^\ast\ & =\ &\left(\begin{array}{cc}
\cos\theta^- & -\sin\theta^- \\
\sin\theta^- & \cos\theta^- \end{array}\right)\; 
\left(\begin{array}{cc}
1 & 0 \\
0 & e^{i\delta^-}\end{array}\right)\; ,\\
{\mathcal V}\ & =\ &\left(\begin{array}{cc}
\cos\theta^+ & -\sin\theta^+ \\
\sin\theta^+ & \cos\theta^+ \end{array}\right)\; 
\left(\begin{array}{cc}
1 & 0 \\
0 & e^{i\delta^+}\end{array}\right)\; ,
\end{eqnarray} 
with 
\begin{eqnarray}
  \label{chmixing}
\delta^- \!&\equiv &\! \arg\left(M_2\Phi_1^{0\ast}
+\lambda S^\ast\Phi_2^{0\ast}\right)\ ,\nonumber\\
\delta^+ \!&\equiv &\! \arg\left(M_2\Phi_2^{0\ast}
+\lambda S \Phi_1^{0\ast}\right)\ ,\nonumber\\
s^- \!&\equiv &\! \sin\theta^-\ =\ \sqrt{m_{\chi_1}^2 - M_2^2 -
\frac{1}{4}\, g_w^2 |\Phi^0_2|^2\over m^2_{\chi_2}-m^2_{\chi_1}}\ ,
\nonumber\\
s^+ \!&\equiv &\! \sin\theta^+\ =\ \sqrt{m_{\chi_1}^2-M_2^2-
\frac{1}{4} g_w^2|\Phi_1^0|^2\over m^2_{\chi_2}-m^2_{\chi_1}}\ ,
\end{eqnarray} 
and $c^\pm \equiv \cos\theta^\pm = \sqrt{1 - (s^\pm)^2}$.
The chargino masses are given by
\begin{eqnarray}
  \label{mchi12}
m^2_{\tilde \chi_{2(1)}} & = & 
{1\over 2}\left[
M_2^2+\lambda^2|S|^2+2g_w^2\left(|\Phi_1^0|^2+|\Phi_2^0|^2\right)
\right.\\
&& \left.+(-)\sqrt{
\left(M_2^2+\lambda^2|S|^2+2g_w^2\left(|\Phi_1^0|^2+|\Phi_2^0|^2\right)
\right)^2
-4|M_2\lambda S-2g_w\Phi_1^0\Phi_2^{0\ast}|^2}\right]\; .
\nonumber
\end{eqnarray}

\end{appendix}

\newpage


\end{document}